**Millimeter-Scale, Atomically Controlled 2D Topological Insulators Revealed by Multimodal Spectroscopy**

Woojoo Lee[1,2], Qiang Gao[1], Yufei Zhao[3], Hui Li[4], Albert Tsui[4], Yichao Zhang[5, §], Yunhe Bai[1], Haoran Lin[1], Khanh Duy Nguyen[1], Gabriele Berruto[1], Gangbin Yan[1], Jianchen Dang[6], Tongyao Wu[6], Hossein Rokni[1], Thomas S. Marchese[1], Ying Shirley Meng[1], Chao-Xing Liu[7], Xiao-Xiao Zhang[6], Chong Liu[1], Pinshane Y. Huang[5], Mark C. Hersam[4], Binghai Yan[3], and Shuolong Yang[1,*]

[1]*Pritzker School of Molecular Engineering, The University of Chicago, Chicago, Illinois 60637, USA*

[2]*Department of Physics, Gachon university, Seongnam 13120, South Korea*

[3]*Department of Condensed Matter Physics, Weizmann Institute of Science, Rehovot 7610001, Israel*

[4]*Department of Materials Science and Engineering, Northwestern University, Evanston, Illinois 60208, USA*

[5]*Department of Materials Science and Engineering, University of Illinois Urbana-Champaign, Urbana, Illinois 61801, USA*

[6]*Department of Physics, University of Florida, Gainesville, FL 32611, USA*

[7]*Department of Physics, Pennsylvania State University, University Park, PA 16802, USA*

[§]Current affiliation: Department of Materials Science and Engineering, University of Maryland, College Park, Maryland 20742, USA

*Corresponding author. Email: yangsl@uchicago.edu

## Abstract

Quantum spin Hall insulators, or synonymously known as 2D topological insulators, are crucial 2D systems hosting topologically protected edge states. The working temperature of this topological quantum phase is dictated by the inverted bandgap. However, the previously identified large-gap 2D topological insulators are either extremely chemically unstable, or cannot be made with atomistic precision over macroscopic scales. Here, we establish two-quintuple-layer $Bi_2Te_3$ and $MnBi_2Te_4/Bi_2Te_3$ heterostructures as atomically controlled, millimeter-scale 2D topological insulators, enabled by precision layer-by-layer growth that yields a carpet-like morphology extending coherently over macroscopic distances. This carpet-like growth mode renders the films amenable to mechanical exfoliation and subsequent wet or dry transfer. Multimodal spectroscopies and microscopies reveal the integer-layer tuned electronic structure of $(Bi_2Te_3)_n$ with excellent agreement to theory. Photon-energy-dependent photoemission and time-resolved photoemission identify band inversion and band dynamics, respectively, while scanning tunneling spectroscopy resolves topological edge states, characteristic of the 2D topological insulator phase. Thickness-and photon-energy-dependent photoemission further validates $MnBi_2Te_4/Bi_2Te_3$ as a robust 2D topological insulator. The large inverted gaps of ~100 meV in $(Bi_2Te_3)_2$ and ~150 meV in $MnBi_2Te_4/Bi_2Te_3$ suggest operation near ambient temperature. These results define a scalable materials platform for next-generation, low-loss quantum and energy-efficient devices.

**Keywords:** Multimodal spectroscopy, layer-by-layer synthesis, topological material, quantum spin Hall insulator, scalable quantum materials platform

**TOC**

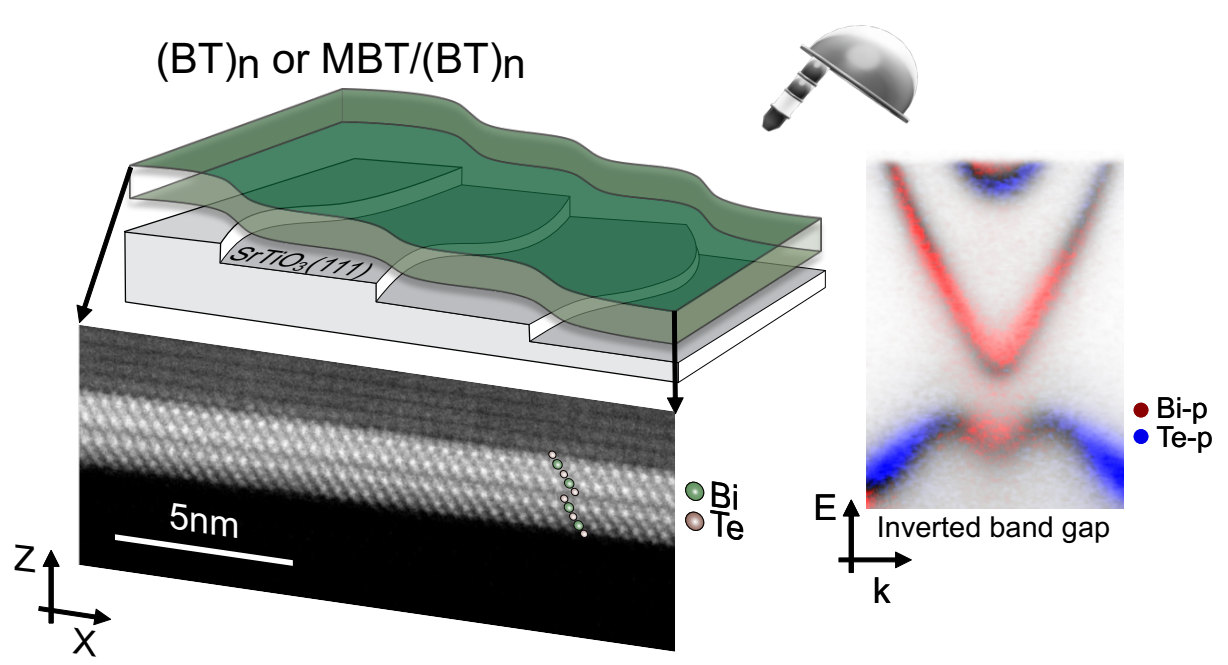

Low-dimensional topological quantum phenomena impose exceptionally stringent requirements on the quality of materials. Crystal symmetry [1], band parity [2], and magnetism [3,4] can drastically vary when the material thickness is off from design by just one functional unit cell, and all of these variations impact the overall topology. For instance, two quintuple-layer $Bi_2Te_3$ (2 QL BT) has long been predicted to be a 2D topological insulator (TI), or a quantum spin Hall insulator (QSHI) [2], where helical edge states lead to a quantized two-terminal resistance of $\frac{h}{2e^2}$. However, the addition of one more QL is expected to render it topologically trivial. Similarly, $MnBi_2Te_4$ (MBT) with an odd number of septuple layers (SLs) behaves as a quantum anomalous Hall insulator (QAHI) with a quantized Hall resistance [4], but the introduction of an extra SL converts the material into an axion insulator with a vanishing Hall resistance [3]. Realizing such ultra-layer-sensitive topological phases requires synthesis that is strictly layer-by-layer — an outstanding challenge for epitaxy and a central bottleneck for scalable topological device architectures.

In this work, we focus on material candidates capable of realizing high-temperature, device-grade 2D TIs. Despite the convention that “2D TIs” and “QSHIs” have been used synonymously [5–8], 2D TIs only exhibit a quantized spin Hall conductance if a spin U(1) symmetry exists and prevents the spin mixing. We will use 2D TIs consistently in this work to emphasize the nontrivial topology. High-temperature, device-grade 2D TIs need to satisfy the following criteria: (i) an inverted gap well exceeding room temperature energy scale, (ii) chemical stability, and (iii) atomically-controlled uniformity across macroscopic distances. Many large-gap 2D TIs, such as Stanene [9,10], Bismuthene [11], and ultra-thin $Na_3Bi$ [12], are extremely chemically unstable. More stable candidates, such as $ZrTe_5$ [13] and 2M-$WSe_2$ [14] are weak 3D TIs and behave as 2D TIs only on individual atomic terraces spanning 10-100 nm, far from device-scale uniformity. In this work, we show that 2 QL BT and MBT/BT hetero-bilayer films, grown by molecular beam epitaxy (MBE), satisfy all three outlined criteria. The large inverted gaps exceeding 100 meV have been predicted by theory [15]. The materials are reasonably chemically robust, signified by the stability of the corresponding bulk compounds. The principal roadblock has been achieving atomically-controlled uniformity across macroscopic distances. Mechanical exfoliation yields BT flakes typically < 20 μm across [16], and MBE growth often produces pyramid-like BT surface morphologies [17–19] with uniform thickness observed only on the 100-nm scale [18,20]. Given the oscillatory topology of ultrathin BT as a function of the number of QLs [2], establishing 2 QL BT and MBT/BT hetero-bilayer films as high-temperature, device-grade 2D TIs requires large-scale, atomically-controlled synthesis combined with multimodal spectroscopic validation of the layer-by-layer topology evolution.

Here, we demonstrate 2 QL BT films and MBT/BT bilayers as millimeter-scale, atomic-layer-controlled 2D TIs with inverted gaps exceeding 100 meV. These ultrathin materials grow coherently across substrate step edges with minimal disruptions, demonstrating a scalable growth method suitable for large-area device integration. Angle-resolved photoemission spectroscopy (ARPES) measurements reveal electronic structures in $(BT)_n$ and MBT/$(BT)_n$ that evolve with integer-layer precision. The layer-by-layer correspondence between the ARPES spectra and the theoretical calculations on BT films reveals the long-predicted transition from a 3D TI to a 2D TI. Time-resolved ARPES (trARPES) uncovers a characteristic $\pi$-phase shift between the conduction band (CB) and valence band (VB) oscillations in 2 QL BT, directly revealing the inverted band character. Scanning tunneling spectroscopy (STS) resolves the topological edge states. Moreover, by converting the top BT QL into an MBT SL, we realize MBT/BT bilayers with even stronger interlayer hybridization and an enlarged inverted gap of 150 meV, stabilizing a 2D TI in the paramagnetic configuration. Finally, these films can be transferred to arbitrary substrates via wet or dry processes. By achieving millimeter-scale uniformity together with > 100 meV inverted gaps, our work opens a realistic materials pathway toward ambient-temperature, wafer-scale topological quantum architectures.

## Results and Discussion

We synthesize $(BT)_n$ and $MBT/(BT)_n$ thin films on $SrTiO_3(111)$ substrates with precisely controlled thicknesses. The uniformity is inspected by atomic force microscopy (AFM), which indicates that the uniformity is maintained at approximately 90% (Figure S1). This allows us to observe discretely varied, layer-dependent electronic structures (Figure S2). In addition to the real-space inspection, micro-ARPES spectra taken within a $1 \times 1$ mm$^2$ region of 2 QL BT confirm that the uniformity extends to millimeter scales (Figure S3); the 6-fold symmetry of Fermi surfaces reveals the high crystallinity of the grown films (Figure S4); the linewidth of the topological surface state (TSS) of 3 QL BT, which is a measure of the electronic scattering rate, is only 20% larger than that of the best cleaved BT crystal (Figure S5). Notably, this uniform morphology applies broadly to $(BT)_n$, $MBT/(BT)_n$, and $(Sb_2Te_3)_n$ films (Figure S1).

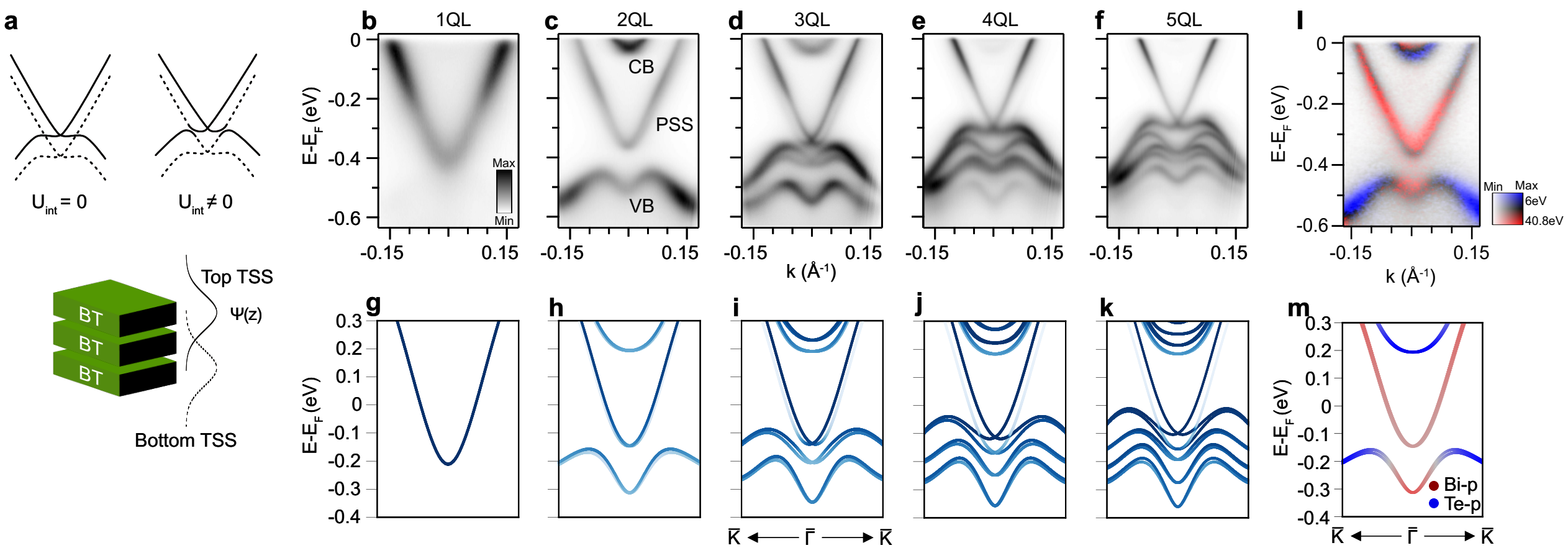

**Figure 1. $Bi_2Te_3$ thin films grown on $SrTiO_3(111)$ substrates and their electronic band structures.** (a) Schematic illustration of the band hybridization between the top (solid line) and bottom (dotted line) topological surface states (TSSs). (b, c)**,** Evolution of band structures for 1-5 quintuple layers (QLs) of $Bi_2Te_3$ films measured by ARPES. CB, VB, and PSS stand for the conduction band, valence band, and pseudo surface state, respectively, in 2 QL $Bi_2Te_3$. (g-k)**,** Theoretically predicted band structures for 1-5 QLs of $Bi_2Te_3$ under the influence of an electric field, $E_\perp = 0.003\ V/Å$. The color saturation indicates the amount of wavefunction projection to the top layer. (l) Dichroism in ARPES spectra for 2 QL $Bi_2Te_3$ derived from two photon energies. (m) Calculated orbital-resolved band structure of 2 QL $Bi_2Te_3$, highlighting band inversion.

ARPES spectra for 1 to 5 QL BT films are presented in Figure 1b-f. We identify three distinct features that attest to the high quality of the films. First, there is an excellent mapping between the intended thicknesses and the observed spectra. Particularly, in the VB region ($E - E_F < -0.3$ eV) the sub-bands resulting from the *c*-axis quantum confinement are clearly distinguished for each thickness. The sub-bands in the CB region are mostly unoccupied precluding a theory-experiment comparison. Theoretically, the number of sub-bands increments linearly as a function of the number of QLs. This progression is in agreement with our experimental data, with the caveat that the calculated detailed band dispersions deviate from experiment presumably due to the absence of quasiparticle interaction terms in the calculation [15,21] (Figure 1g-k).

Moreover, secondary Dirac cones are observed at 77 meV and 118 meV below the primary ones in the 3 QL and 4 QL BT films, respectively (Figure 1d-e). We plot the corresponding ARPES spectra using nonlinear color scales in Figure S6 for better visualization of the secondary Dirac cones. The spectral weight ratio between the secondary and primary Dirac cones is 0.41, 0.14 and 0.02 in the 3 QL, 4 QL and 5 QL films, respectively, using momentum distribution curves (MDCs) at 0.1 eV above the primary Dirac point (Figure S6). Both the binding energy progression and the intensity evolution of the second Dirac cone support the attribution of this feature to the TSS from the bottom surface (bottom-TSS), as we argue below. An electric field, originating from the polar $SrTiO_3$ surface [22] or band bending in BT films [23], disrupts the inversion symmetry. This electric field induces the energy separation between the top- and bottom-TSSs, with the separation increasing as a function of thicknesses. Moreover, the photoelectrons originating from the bottom-TSS experience a longer escape path compared to those from the top-TSS, resulting in a weaker photoelectron intensity. As a result, the bottom-TSS is barely identifiable in the BT films thicker than 4 QL. These observations are fully reproduced by first-principles calculations (Figure 1g-k).

Lastly, we observe the anti-crossing gaps between the top- and bottom-TSSs. In the 3 QL film, a gap of approximately 30 meV is extracted via Gaussian fitting (Figure S7), whereas in the 4 QL film the gap becomes smaller than our experimental resolution, which suggests that starting from 4 QL the two TSSs become localized on each side of the BT film and exhibit minimal interactions. Hence, the 3D TI phase of BT starts at the thickness of 4 QL.

The 2 QL BT film enters a distinct 2D phase. Our calculations yield inverted band characters across the bandgap (Figure 1m) and topologically nontrivial edge states within the gap (Figure S8), elucidating the 2D TI phase. The general evolution of the hybridization between the top and bottom TSSs is consistent with a previous study on ultrathin $Bi_2Se_3$,[23] yet its conclusion regarding the 2D TI phase was based on band dispersion fitting, with no direct experimental evidence for the nontrivial topology. In a previous study reporting layer-by-layer growth of BT,[24] the bottom surface states were not resolved and the spectra for 2 QL and 3 QL BT were indistinguishable possibly due to thickness mixing. In our study on 2 QL BT, however, we will provide multiple independent experimental signatures for the 2D TI phase. The inverted band characters can be revealed in photon energy dependent ARPES studies [25]. The Bi 6*p* and Te 5*p* orbitals, which together constitute the CB and VB near $E_F$, have different ionization cross-sections at varying photon energies [26] (Supplementary Note 1). This enables an investigation of the orbital origins of the bands. By taking the spectral difference between the ARPES data acquired using 6 eV and 40.8 eV photons with a proper scaling factor (Supplementary Note 1), we obtain a dichroism map (Figure 1l) in agreement with the calculated orbital character map (Figure 1m).

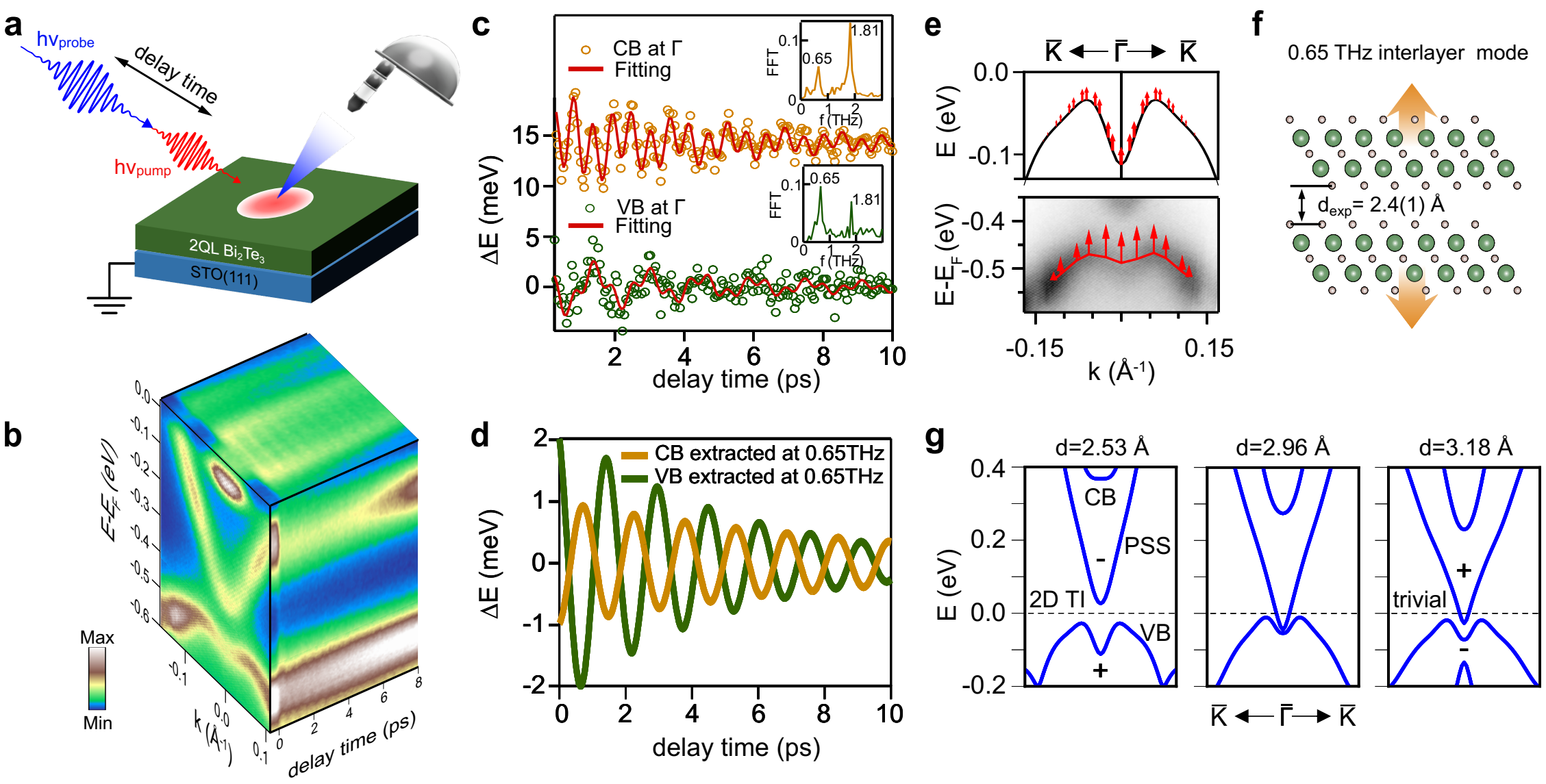

**Figure 2. Inverted bandgap in 2 QL $Bi_2Te_3$ modulated by a coherent interlayer phonon mode.** (a) Schematic of the time-resolved ARPES experiment. (b) Three-dimensional mapping of photoemission intensity as a function of energy, momentum, and pump–probe delays. (c) Energy oscillations of the conduction and valence bands (CB and VB) at $\bar{\Gamma}$, tracked by time-resolved ARPES. The assignments of CB and VB are marked in Figure 1c. The oscillations are fitted using exponentially decaying double-cosine functions with 0.65 THz and 1.81 THz frequencies for inter- and intra-layer modes. The insets display the Fast Fourier Transforms of the energy oscillations. A narrow momentum window of ±0.01 $Å^{-1}$ centered at the Γ point was used for the energy distribution curve (EDC) analysis. (d) Extracted energy oscillations of the 0.65 THz mode from Panel (c). (e) Top: Calculated energy oscillation amplitudes on VB induced by the interlayer phonon mode using frozen-phonon density functional theory. Bottom: Momentum dependent energy oscillation amplitudes on the M-shaped VB illustrated by red arrows. (f) Illustration of the 0.65 THz interlayer phonon mode. (g) Calculated change of the electronic band structure toward a trivial band topology driven by the systematically enlarged interlayer distance.

To further substantiate the inverted band characters, we conduct trARPES experiments on 2 QL BT and investigate the electronic structure changes in response to coherent lattice distortions (Figure 2a, 2b). Specifically, our focus is on the interlayer phonon mode (Figure 2f), which modulates the interlayer van der Waals (vdW) distance. In 2 QL BT, we observe two modes at 0.65 THz and 1.81 THz (Figure 2c). The 0.65 THz mode is rapidly softened as the thickness increases, which is a signature of the interlayer mode [27] (Figure S11). The frequency of the 1.81 THz mode exhibits a weak thickness dependence, which is consistent with the attribution to an intralayer $A_{1g}$ phonon [28–30]. By modulating the vdW distance using the interlayer phonon mode, we provide time-domain evidence for the topologically nontrivial phase of 2 QL BT. We emphasize that our experiment is conducted in the perturbative regime where the system is far from a transient topological phase transition. However, by studying the perturbative changes in the band structure we reveal the topological nature. Importantly, our STEM measurement determines the equilibrium vdW distance to be 2.4 ± 0.1 Å (Figure S12). In comparison with the theoretical upper limit of the vdW distance for stabilizing the topologically nontrivial phase (2.96 Å, Figure 2g), the STEM measurement putatively puts the 2 QL BT system in the 2D TI phase. Building on this premise, frozen-phonon calculations – executed without the electric field effect for simplicity, given that the $E_{\perp} = 0.003\ V/\text{Å}$ (see Figure 1) is much smaller than the electric field required for topological phase transition [31] – provide three falsifiable predictions when the vdW distance is modulated (Figure 2g). We compare these three predictions with experimental observations, thus examining the premise.

First, the central dispersion dip on the M-shaped VB becomes less pronounced as the vdW distance increases, driving the system from the 2D TI regime toward the topological transition. Experimentally, the momentum dependent oscillation amplitudes on the M-shaped VB are substantially larger near $\bar{\Gamma}$, suggesting that this lattice distortion directly modulates the central dispersion dip, in agreement with theory (Figure 2e). Second, our calculations show that the CB and VB will be shifted in opposite directions only in the topologically nontrivial phase. This effect is shown theoretically as we enlarge the interlayer distance from 2.53 to 2.96 Å (Figure 2g). To examine this, we fit the energy oscillations of the CB and VB using two exponentially decaying cosine functions, and extract the oscillatory components at 0.65 THz. The energy oscillations of the CB at 0.65 THz exhibit a π phase shift with respect to those of the VB (Figure 2d and Figure S13), which verifies the prediction. We note the correspondence between the topologically nontrivial phase and the $\pi$ phase shift between CB and VB oscillations is material specific, and does not generally apply to other TI materials such as $Bi_2Se_3$ [27]. Third, an additional VB below the M-shaped band is predicted to emerge only in the trivial phase (Figure 2g). This feature is not observed in any of our ARPES experiments (Figure S10). Taken together, these three independent observations show excellent agreement with the theoretical predictions and support the proposed inverted electronic structure consistent with a 2D topological insulator phase in 2 QL BT.

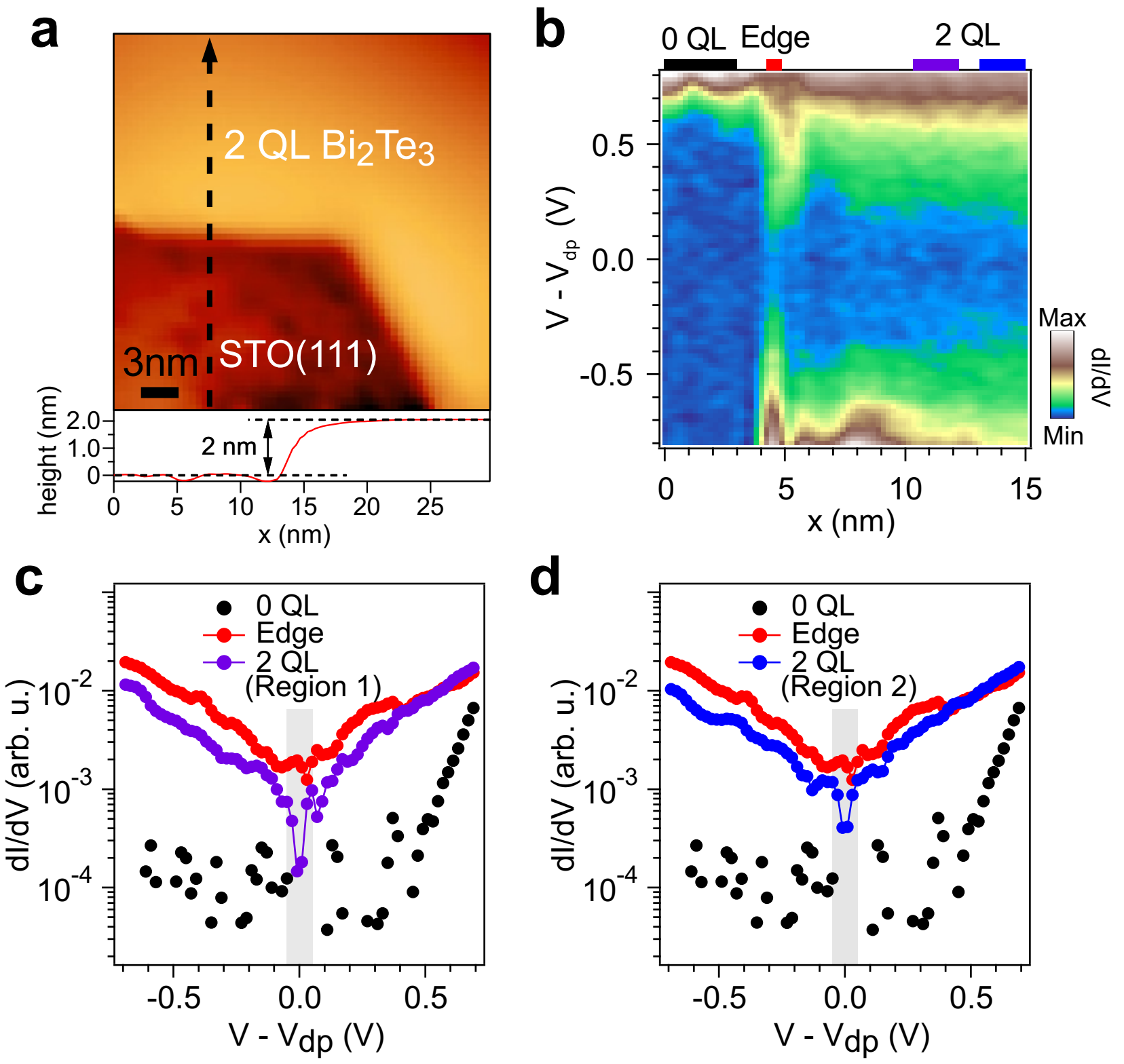

**Figure 3. Local detection of topological edge states using scanning tunneling spectroscopy (STS).** (a) STM image of a boundary between 2 QL $Bi_2Te_3$ and bare $SrTiO_3$ substrate, $V_{bias}$ = 1 V, $I$ = 0.4 nA. The height profile taken along the black dashed line reveals a step height of 2 nm, corresponding to the thickness of 2 QL $Bi_2Te_3$. (b) Color map of differential conductance (dI/dV) spectra across the boundary between 0

QL and 2 QL regions using STS. The scan direction is indicated in panel (a) with a black dashed arrow. DP: Dirac point. (c, d) dI/dV spectra for 0 QL, 2 QL and the edge, integrated in the spatial regions marked by colored symbols on top of panel (b). After subtracting the background value - obtained by averaging over $SrTiO_3$'s spectrum in the gapped bias energy range - from all dI/dV spectra, some data points of the $SrTiO_3$ spectrum fall below zero, hence not shown on the log scale. $V_{dp}$ is practically defined by the dips of the 2 QL spectra. No vertical offset is applied to panels (c) and (d).

The bulk-edge correspondence in a 2D TI system gives rise to topologically protected 1D edge states. We examine these states near the boundary between the topologically trivial 0 QL and non-trivial 2 QL regions (Figure 3a) using scanning tunneling spectroscopy (STS). The differential conductance (dI/dV) color map across the 0 QL/2 QL edge explicitly reveals changes of the electronic structure (Figure 3b), transitioning from a bandgap > 1.2 eV associated with the $SrTiO_3$ substrate to a 0.1 eV bandgap characteristic of 2 QL BT (Figure 1c). By inspecting the dI/dV curves on a logarithmic scale in two separate 2-QL regions (Figure 3c, 3d), we repeatedly observe spectral dips, where the spectral intensity approaches the noise floor defined by the gapped region of the STO spectrum. However, quantitatively extracting the gap size is challenging due to position-dependent fluctuations in the spectral intensity. We speculate that this may arise from local impurities within the BT film or even from the STO substrate beneath it. Nevertheless, judging from the turning points that define the spectral dips, our dI/dV curves in both Regions 1 and 2 of the 2-QL BT are consistent with a ~0.1 eV gap (fitting details in Figure S14), as also resolved by ARPES.

Especially, at the edge we observe an enhanced dI/dV intensity compared to that from the 2 QL region. This enhanced dI/dV intensity occurs at bias voltages corresponding to the energy gap of 2 QL BT, indicative of in-gap 1D edge states (Figure 3c, 3d). The spectral intensity enhancement is also seen in an extended bias voltage range outside the inverted gap, in agreement with observations on other QSHI systems [8,12,32]. Moreover, absence of the topological edge states is confirmed at a 0 QL/1 QL edge (Figure S15), as both sides of this edge are topologically trivial.

Notably, experimentally verifying the topological nature of these states is challenging, since they can persist even under applied magnetic fields [32,33]. Despite these challenges, the 2D TI phase in 2 QL BT is established by combining multiple independent experimental signatures: the layer-by-layer correspondence between our experimental and theoretical electronic structures for the 1-5 QL BT films, the orbital origins and transient band dynamics of 2 QL BT, as well as the observation of the 1D edge states.

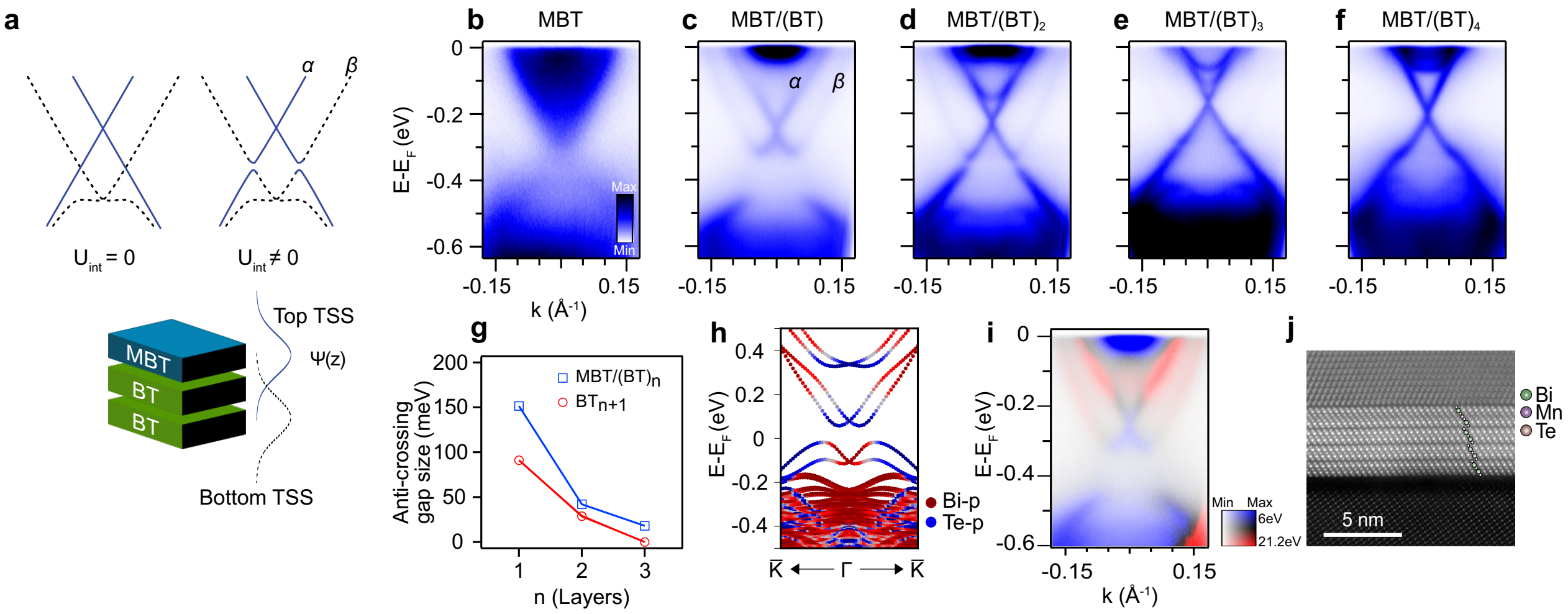

**Figure 4. Creation of $MnBi_2Te_4/(Bi_2Te_3)_n$ heterostructures and their electronic band structures.** (a) Schematic illustration of the band hybridization between the top (solid line) and bottom (dotted line) topological surface states (TSSs). $\alpha$ and $\beta$ denote the top and bottom TSSs, respectively. (b-f) Evolution of band structures measured by ARPES for $MnBi_2Te_4/(Bi_2Te_3)_n$ with systematically incremented $n$ ($n = 0, 1, 2, 3, 4$). (g) Measurement of the anti-crossing gaps in $(Bi_2Te_3)_{n+1}$ and $MnBi_2Te_4/(Bi_2Te_3)_n$ films. The error bars are approximately 1 meV, which is smaller than the symbol size and therefore not visually distinguishable. (h) Orbital-resolved band structure of the $MnBi_2Te_4/Bi_2Te_3$ bilayer system in the paramagnetic phase, highlighting band inversion. The red and blue colors represent the contributions from the Bi-*p* and Te-*p* orbitals. (i) Dichroism in ARPES spectra derived from 6 eV and 21.2 eV highlighting the contrast between the Bi and Te orbital characters. (j) Cross-sectional STEM image of the $n = 3$ heterostructure viewed along the [ $\bar{1}$ 1 0] direction of $SrTiO_3(111)$, capped by Te overlayers.

The systematic characterization of 2 QL BT demonstrates that layer and photon-energy dependent ARPES can identify 2D TIs. This enables us to use ARPES to efficiently screen additional 2D TI materials, among which is the $MBT/(BT)_n$ system. The electronic band structures of these heterostructures exhibit striking changes compared to the original BT counterparts (Figure 4b-f). We designate each heterostructure as $MBT/(BT)_n$ with varying $n$ ($n = 0, 1, 2, 3, 4$) representing the number of BT layers. The ARPES spectrum of monolayer MBT ($n = 0$) shows diffuse features representing the CB and VB, and is consistent with previous results on MBE-grown monolayer MBT [34]. In the $n = 1$ heterostructure (Figure 4c), an X-shaped band emerges near $\bar{\Gamma}$, denoted as the $\alpha$ band. The band dispersion is disrupted by a 150 meV gap centered around

-0.4 eV. A faint electronic band ($\beta$) with a linear dispersion and Fermi momenta at $\pm 0.15$ Å$^{-1}$ connects with the $\alpha$ band near -0.3 eV, giving the impression of a Rashba-like band splitting. However, the linear dispersions of both bands indicate that they cannot be understood as simple near-free-electron-like Rashba states. We argue that the large gap disrupting the $\alpha$ band dispersion is an anti-crossing gap due to the hybridization between the $\alpha$ and $\beta$ bands (Figure 4a). In the $n = 2$ heterostructure, the $\alpha$ band becomes more prominent, and the anti-crossing gap diminishes to 42 meV (Figure 4d). Given the decreasing anti-crossing gap and the decreasing intensity of the $\beta$ band as a function of the film thickness, we attribute the $\beta$ band to the bottom-TSS with its Dirac point around -0.5 eV (Figure 4a). Correspondingly, the $\alpha$ band is assigned to the top-TSS (Supplementary Note 2). The evolution of the hybridization between the top- and bottom-TSSs is similar to that in pure BT films, yet the energy separation between them is larger due to the strong interlayer electron mixing at the MBT/BT interface (Figure 4g) [35].

The photon energy dependent ARPES dichroism (Figure 4i) agrees with our first-principles calculation of the orbital characters (Figure 4h), elucidating that the MBT/BT bilayer heterostructure in the paramagnetic (PM) phase has an inverted bandgap and is in the 2D TI phase. Here we use 6 eV and 21.2 eV photons, since the photoemission cross-section using 40.8 eV on MBT/BT bilayer films is extremely low which indicates the lack of final states (Figure S17). Experimentally, the inverted bandgap in an MBT/BT bilayer is 150 meV, which is even larger than the counterpart in 2 QL BT (Figure S18 and S19). This gap persists at temperatures both below and above the ferromagnetic critical temperature of ~20 K as determined by reflective magnetic circular dichroism (RMCD, Supplementary Note 3). Nevertheless, prior theoretical studies have shown that exchange coupling in MBT-based heterostructures can modify the topological phase depending on its strength[35,36]. A thorough analysis of the impact of ferromagnetism on the electronic structure is available in Supplementary Note 4. In the main text, we focus on the paramagnetic phase which is particularly pertinent to the study of 2D TIs. The 150 meV inverted bandgap originates from the mixing of outer shell electrons of Bi and Te atoms near the interface, leading to a charge transfer between the MBT and BT layers and an enhanced interlayer coupling [35]. This mechanism is manifested in the 180 meV splitting between the MBT- and BT-derived Dirac cones (Figure 4c).

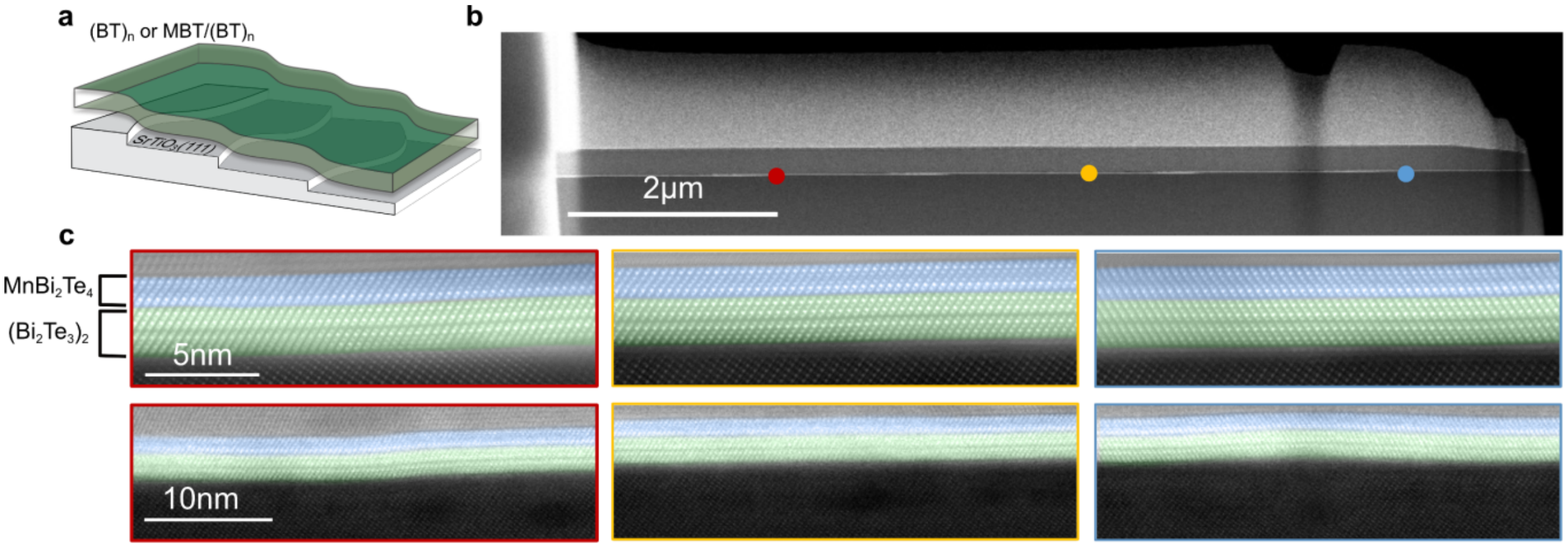

**Figure 5. Millimeter-scale uniformity of ultrathin $(Bi_2Te_3)_n$ and $MnBi_2Te_4/(Bi_2Te_3)_n$ films grown in a carpet mode.** (a) Illustration of the carpet mode. (b) Cross-sectional scanning transmission electron microscopy (STEM) image taken from a $MnBi_2Te_4/(Bi_2Te_3)_2$ lamella, capped by Te and then Pt. (c) Atomic-resolution STEM images taken at color-coded points in panel (b) with approximately 3 $\mu m$ intervals, viewed along the $[\bar{1}10]$ direction of $SrTiO_3$. The high-resolution STEM images demonstrate atomic-level uniformity across a 6 $\mu$m specimen. The blue and green false colors on the STEM images indicate the $MnBi_2Te_4$ and $Bi_2Te_3$ layers.

To understand the atomic origin of our layer-by-layer controlled electronic structures, we perform STEM measurements on $MBT/(BT)_2$ (Figure 5 and S26) and $(BT)_2$ (Figure S27). A series of cross-sectional STEM images, taken with approximately 3 μm intervals within an $MBT/(BT)_2$ lamella, suggests that the film extends continuously across the substrate at both micro- and nano-meter scales, undisrupted by the step edges, ridges, or other non-uniformities on the substrate (Figure 5). This unique carpet growth mode is likely enabled by the minimization of free energy when the single-layer thickness (~1 nm) is much larger than the substrate step height (0.22 nm)[21] (Supplementary Note 5), yet we emphasize that the 3×3 surface reconstruction on the $SrTiO_3(111)$ substrate and a careful tuning of the substrate temperature are also crucial (Supplementary Note 6).

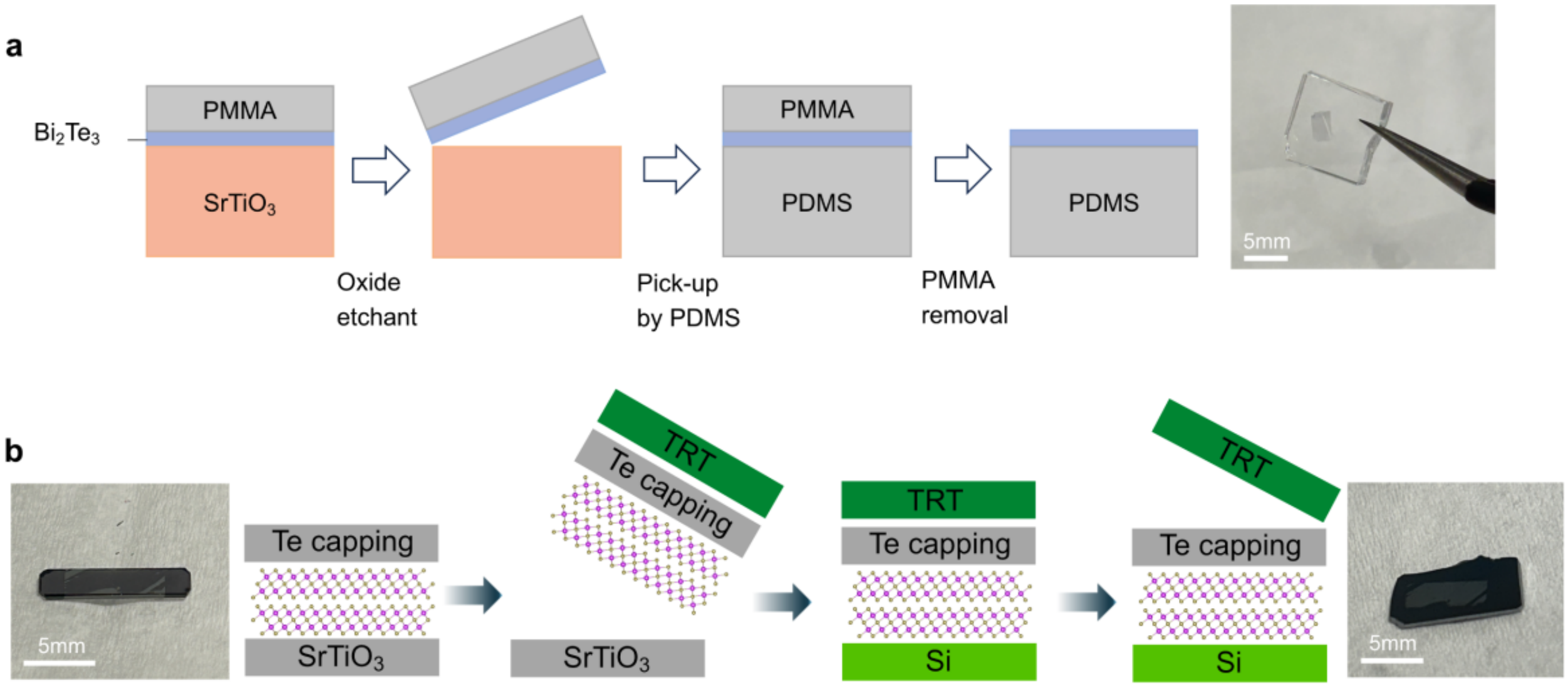

**Figure 6. Release and transfer of MBE-grown $Bi_2Te_3$ films.** (a) Schematic illustration of a wet-transfer process. A poly(methyl methacrylate) (PMMA) layer is used to keep the flatness of the 2 QL $Bi_2Te_3$ film, while a buffered oxide etchant removes the interfacial layer of the $SrTiO_3$ substrate, enabling the subsequent transfer to a PDMS gel. (b) Schematic illustration of a dry-transfer process. A thermal release tape (TRT) enables direct exfoliation of the Te-capped film from the $SrTiO_3$ substrate and transfer onto a Si substrate.

Importantly, the carpet-mode growth reduces the film–substrate adhesion, allowing both a wet-transfer via selective substrate etching of a 2×3 $mm^2$ 2 QL BT film onto a polydimethysiloxane (PDMS) gel and a dry-transfer via direct film liberation onto arbitrary substrates (Figure 6). The successful transfer of such millimeter-scale thin films [37] enables future opportunities to build complex topological systems by programmed stacking and twisting of 2D TIs [38–41]. This transferability distinguishes our millimeter-scale 2D TI membranes from the previously discovered 2D TI platforms with large inverted gaps, such as Bismuthene[11] and ultra-thin $Na_3Bi$[12]. For instance, one could imagine that placing a 2-nm thick 2D TI membrane on a stretchable and bendable polymer will facilitate the development of a flexible topological field effect transistor (FET) which utilizes the TI membrane as the "channel" material [42]. The chemical potential will need to be tuned into the inverted gap either by Sb-doping [43] or by electrical gating [44]. We emphasize that the coherence length for quantized charge transport can be limited to a few micrometers due to spin dephasing [45], but the prospect of millimeter-scale membrane transfer offers an expedited pathway to fabricate a large array of interconnected topological devices [46].

## Conclusion

By revealing the electronic structures of ultrathin $(BT)_n$ and $MBT/(BT)_n$ films, we have, for the first time to our knowledge, demonstrated a clear agreement between ARPES measurements and first-principles calculations on these materials, and thus tremendous potentials for topology engineering in the thickness range of 2~4 nm [35,36]. Our work has profound implications for the MBE growth of ultrathin topological materials. Notably, successful implementations of low-dimensional topological physics using MBE, such as the QAH effect in $Cr_{0.15}(Bi_{0.1}Sb_{0.9})_{1.85}Te_3$ [47] and the QSH effect in HgTe/(Hg,Cd)Te quantum wells [7], were seminal milestones in the field which leveraged the fact that these systems were robust against thickness fluctuations of 1~2 layers. In this work, on the other hand, the successful creation of the millimeter-scale $(BT)_n$ films and $MBT/(BT)_n$ heterostructures offers exciting prospects for band structure engineering and the exploration of a variety of topological quantum phases which do require a precise layer number control [35,36,48–50]. This represents substantial advances over previous results in the literature on ultrathin BT [24] and MBT [51,52] films, where the limited thickness uniformity prevented the visualization of layer-specific electronic structures. Remarkably, the inverted gaps for 2 QL BT and MBT/BT bilayer films are 100 meV and 150 meV, respectively, which are more than a factor of two larger than that in HgTe/(Hg,Cd)Te quantum wells (40 meV) [7] and 1T'-$WTe_2$ (45 meV) [8]. Considering that a QSHI was realized at 100 K in 1T'-$WTe_2$ [53], 2 QL BT and MBT/BT bilayers are promising platforms to realize QSHIs potentially near ambient temperatures.

## Methods

### Sample growth

The high-quality $Bi_2Te_3$ films were prepared by co-evaporating high-purity sources of Bi (99.999%) and Te (99.9999%) onto Nb-doped $SrTiO_3(111)$ substrates at a temperature of 240°C within an ultra-high vacuum (UHV) molecular beam epitaxy (MBE) chamber. The films were then transferred *in situ* to an angle-resolved photoemission spectroscopy (ARPES) chamber for precise characterization, and subsequently transferred back to the MBE chamber. To convert the top BT layer to the MBT layer, the sample temperature was raised to 270°C, and the co-evaporation of Mn (99.9998%) and Te (99.9999%) sources was carried out. For more detailed information, refer to the Supporting Information.

## ARPES measurements

All ARPES measurements, unless otherwise stated, were carried out at 13 K on the Multi-Resolution Photoemission Spectroscopy (MRPES) platform established at the University of Chicago (Figure S30).[54] Ultrahigh resolution 6 eV μARPES measurements were performed with a spatial resolution <10 μm and an energy resolution < 4 meV using 205 nm probe pulses with 80 MHz repetition rate. Additional ARPES measurements using 21.2 eV (He 1$\alpha$) and 40.8 eV (He 2$\alpha$) photons were also performed with an energy resolution < 6 meV. Ultrafast trARPES measurements were performed with an energy resolution of 17 meV and a time resolution of 115 fs at 200 kHz repetition rate. Pump and probe wavelengths were 800 nm and 206 nm, respectively. The incident fluence was 63 $\mu J/cm^2$.

## RMCD measurements

For the RMCD measurements, a HeNe laser of 1.96 eV was used. Measurements were taken inside a closed-cycle optical cryostat (attDry 1000) with a base temperature of 3.6 K and a superconducting solenoid magnet up to 9 T. The laser was modulated at 50 kHz between the left and right circular polarization using a photoelastic modulator (PEM) and focused onto the sample with a 50× objective. The reflected light was collected and focused onto an amplified photodiode. The RMCD was determined as the ratio of the ac component of the photodiode signal measured by a lock-in amplifier at the polarization modulation frequency and the dc component of the photodiode signal measured by a voltmeter.

## STEM measurements

- $MnBi_2Te_4$/ $Bi_2Te_3$ heterostructure -

Cross-sectional lamellae of Te capped $MnBi_2Te_4/(Bi_2Te_3)_n$ were prepared using focused ion beam milling (Thermo Fisher Helios NanoLab 600i) along [$\bar{1}$10] crystallographic direction of $SrTiO_3$ substrate to access the [$\bar{1}$100] zone axis of the $Bi_2Te_3$ layers. The samples were imaged with a Themis Z aberration-corrected scanning transmission electron microscope (Thermo Fisher Scientific), operated at 300 kV. The aberration-corrected ADF-STEM images were acquired using a high-angle annular dark field detector at a convergence angle of 25.2 mrad. Each image was acquired as ten sequential frames (5 μs/pixel) in the same region and frame-averaged using drift corrected frame imaging in Velox software to minimize distortion from sample drift.

- $Bi_2Te_3$ film -

The $Bi_2Te_3$ lamellae were prepared using a focused ion beam system (Thermo Fisher scientific Helios 5CX) at the University of Illinois at Chicago along [$\bar{1}$10] crystallographic direction of $SrTiO_3$ substrate to access the [$\bar{1}$100] zone axis of the $Bi_2Te_3$ layers. STEM images were acquired using the aberration-corrected JEOL ARM200CF at the University of Illinois at Chicago. A cold field emission source operated at 200 kV was equipped. The high-angle annular dark-field (HAADF) detector angle was 90-270 mrad to acquire Z contrast images.

## STM/STS measurements

All STM/STS measurements were performed with a Scienta Omicron LT STM at ~4 K. Samples were transferred from the MRPES at the University of Chicago to the STM at Northwestern University with a vacuum suitcase (base pressure ~ $10^{-10}$ mbar) to protect the surface. The PtIr tip was pre-conditioned on a clean Ag(111) surface. The d*I*/d*V* spectra were taken by a SR850 lock-in amplifier with 5 mV bias modulation and 820 Hz frequency. The spectroscopic mapping was conducted after repeated topographic scans in the interested field-of-view to reduce the drift.

**DFT calculations**

Our first-principles calculations were performed with the Vienna Ab initio Simulation Package (VASP) using the projector-augmented wave method[55,55,56]. The exchange-correlation functional based on generalized gradient approximation (GGA) parameterized by Perdew-Burke-Ernzerhof (PBE) was adopted[57]. To treat the correlation effect of localized 3d electrons of Mn, the DFT+U method by Dudarev et al. was employed with $U_{eff}$ = 5 eV[58]. The kinetic energy cutoff of plane-wave basis was set to 400 eV. Van der Waals corrections of the DFT-D3 method were considered[59]. We adopted the Monkhorst-Pack k-point meshes of 10×10×4 and 10×10×1 for the calculations of $Bi_2Te_3$ bulk and $MnBi_2Te_4$-$Bi_2Te_3$ slabs, respectively. The maximally localized Wannier functions of Bi-6p and Te-5p orbitals were constructed by WANNIER90 package[60]. Based on the bulk tight-binding parameters of $Bi_2Te_3$, we constructed the slab model for few layers and evaluated their band evolution with an electric field $\epsilon = 0.003$ V/Å. To mimic the paramagnetic (PM) band structure of $MnBi_2Te_4$-$Bi_2Te_3$ slab, the "special quasi-random structures" approach was implemented with disordered local moments in a 4×4 supercell[61]. Then by fitting a 2 quintuple layer (QL) $Bi_2Te_3$ Hamiltonian with $\epsilon = 0.02$ V/Å, we obtained the edge green function for this effective PM SL-QL slab.

**Supporting Information:** Additional microscopic and spectroscopic measurements on millimeter-scale carpets of $MnBi_2Te_4/(Bi_2Te_3)_n$ and $(Bi_2Te_3)_n$; Additional calculations for $MnBi_2Te_4/(Bi_2Te_3)_n$ and $(Bi_2Te_3)_n$; Normalization of photoionization cross-sections; Band assignments of bottom and top topological surface states; Ferromagnetic critical temperature; Magnetic gap investigation below the magnetic transition temperature; Step-climbing growth mode; Precisely controlled MBE growth of $MnBi_2Te_4/(Bi_2Te_3)_n$ and $(Bi_2Te_3)_n$; Experimental setup of the integrated MBE-trARPES system.

**Acknowledgments**

**Funding:** The molecular beam epitaxy growth, angle-resolved photoemission spectroscopy (ARPES) characterization, and scanning transmission electron microscopy (STEM) on $Bi_2Te_3$ films were supported by DOE Basic Energy Sciences under Grant No. DE-SC0023317. The substrate preparation and atomic force microscopy were partially supported by National Science Foundation under Award No. DMR-2011854. The time-resolved ARPES work was supported by DOE Basic Energy Sciences under Grant No. DE-SC0022960. The STM work was supported by the National Science Foundation Materials Research Science and Engineering Center at Northwestern University (NSF DMR-2308691). The preparation of $Bi_2Te_3$ films for STEM imaging made use of the ThermoFisher Helios 5CX FIB instrument in the Electron Microscopy Core of the University of Illinois at Chicago's Research Resources Center, which received support from the University of Illinois at Chicago, Northwestern University, and ARO (W911NF2110052). The STEM imaging on $MnBi_2Te_4/(Bi_2Te_3)_n$ was supported by PECASE under Grant No. AF FA9550-20-1-0302 and carried out in the Materials Research Laboratory Central Facilities at the University of Illinois Urbana-Champaign. The authors acknowledge the use of facilities and instrumentation supported by NSF through the University of Illinois at Urbana-Champaign Materials Research Science and Engineering Center DMR-2309037.

**Author contributions:** W.L. and Q.G. grew all the films, and carried out the ARPES and trARPES measurements under the supervision from S.-L.Y., and with assistance by K.D.N., Haoran L., Y.B., and G.B.. Yichao Z. and P.Y.H. took STEM images for $MnBi_2Te_4/(Bi_2Te_3)_n$. G.Y., T. S.M., Y.S.M. and C.L. took STEM images for $(Bi_2Te_3)_n$. J.D., T.W. and X.-X.Z. conducted RMCD measurements. Yufei Z. and B.Y. carried out first-principles calculations with input from C.X.L.. K.D.N., Haoran L. and H.R. performed AFM measurements. H.L., A.T. and M.C.H. conducted STM measurements. W.L. and S.-L.Y. wrote the manuscript with input from all co-authors.

**Competing interests:** The authors declare that they have no competing interests.

**Data and materials availability:**
The data that support the findings of this study are available from the corresponding author upon reasonable request. Data for the main figures are deposited in Zenodo and can be accessed at https://zenodo.org/records/17567720?preview=1&token=eyJhbGciOiJIUzUxMiJ9.eyJpZCI6IjkzNDgyNGM1LTI5YjctNDA0OS05ODE5LTMzOTk1NzNiNjUzMiIsImRhdGEiOnt9LCJyYW5kb20iOiJiNGI2M2Q2NGFkNDFkNjViMGE0NmVmNjgzZGRkYzkzMCJ9.H-OF1ir1LE7BqHYM5zg-TAYa4YptoF8m97yHmXw2nVGSZz1Bre65CfFkxPqbGe098SUpcMdPzmqG9ZyIzWTKcw
(This temporary link is provided for peer review and will be replaced with a DOI upon publication.)